\documentclass[11pt,twoside]{article}
\usepackage{asp2010}

\resetcounters

\begin{document}

\title{Teaching Charge Coupled Devices Using Models as Part of the Engineering Design Process at Maui Community College}
\author{Isar Mostafanezhad$^1$, Johnny Tam$^2$, Ciril Rozic$^1$, David M. Harrington$^3$, Bradley A. Jacobs$^3$, Ryan Swindle$^3$, and Elisabeth Reader$^4$ 
\affil{$^1$Department of Electrical Engineering, University of Hawai'i, Honolulu, HI 96822}
\affil{$^2$Joint Graduate Group in Bioengineering, University of California, Berkeley, CA 94720}
\affil{$^3$Institute for Astronomy, University of Hawai'i, Honolulu, HI 96822}
\affil{$^4$University of Hawai'i, Maui College, Kahului, HI, 96732}}

\begin{abstract}
The CCD Modeling Activity was designed to supplement the curriculum of the Electrical and Computing Engineering Technology program at the Maui Community College. The activity was designed to help learners understand how a Charge Coupled Device (CCD) works.  A team of visiting graduate students was invited to teach an activity through the Teaching and Curriculum Collaborative (TeCC) as part of the Center for Adaptive Optics/Institute for Science and Engineer Educators Professional Development Program. One of the primary goals was to have students gain an understanding of the function of a CCD by constructing a model representing the CCD readout process. In this paper we discuss the design and implementation of the activity and the challenges we faced.
\end{abstract}

\section{Introduction}
The Professional Development Program (PDP) is a training program now run out of the Institute for Science and Engineer Educators (ISEE) that teaches research-based educational techniques to the next generation of educators while simultaneously satisfying goals related to workforce development. As part of this program, participants receive training on how to engage a diverse range of learners and learning styles. In collaboration with the PDP, Maui Community College (MCC; currently known as University of Hawaii Maui College) has invited these instructors into their classrooms to teach newly designed activities. This Teaching and Curriculum Collaborative program (TeCC) brings groups of instructors to MCC to teach in the Electrical and Computing Engineering Technology program (ECET). Examples of other TeCC activities can be found in Morzinski et al.~(this volume).

The activities designed as part of this collaboration focus on an inquiry-based approach to learning, that is, a technique of learning motivated by questions which learners investigate \citep[see][for a discussion]{PDPdesc}.  It attempts to engage learners in activities that mirror the research practices of professional scientists and engineers.  The CCD activity described here is designed around this principle, beginning with an introduction to engineering models and an activity to force students to think about what concepts define a good (or bad) model.  We then allow students to explore the design considerations for CCDs, using very simple materials, to try and remove any technical preconceptions they may have, and to reduce the complexity of the inquiry process.  This simplicity in design seems to work well, as the students are quickly able to use known materials to explore their own designs. We hope that this activity helps learners be able to think more critically about the functions and limitations of advanced technology, using an example that students in the ECET program are likely to encounter in their future careers, namely CCDs.

The ECET program is designed to train future engineering technicians and technologists in a two-year program focused on engineering applications relevant to the Hawaii high-tech industry. The CCD modeling activity was designed to fit in to a course on Instrumentation for Engineering Technology, known as ETRO 102. The course covers the fundamental principles and applications of optics, electronics, engineering and computer software in a variety of disciplines and includes emphasis on data collection, imaging, image processing and Hawaii high-tech industry applications.  The topics covered include geometric optics and images, the nature of digital images and CCDs, color and light, filters, telescopes, remote sensing instrumentation and atmospheric propagation of light.  In a typical ETRO 102 classroom, there are 15 to 20 students made up of mostly first and second year ECET program participants. MCC has a high fraction of minority students, a significant fraction of which have English as a second language. This is reflected in the diverse cultural backgrounds of students in the ECET program.

Three to four PDP participants designed and taught the CCD activity as part of ETRO 102 in the Fall of 2008 and 2009.  The participants instructed 24 (in 2008) and 16 (in 2009) students enrolled in ETRO 102: Instrumentation. The activity took place over the course of one week in two lab sessions lasting 2.5 hours each. The primary content goal of the activity was to have learners understand how the readout process of a CCD works. Secondary content goals included understanding CCD components (e.g. pixels, arrays, counters), applications, and limitations (e.g. speed, efficiency, noise). The primary process goal was to have learners develop an understanding of modeling, and how modeling fits into the overall engineering design process \citep[see][]{masseng}. The use of blueprinting, prototyping, validation, communication, and brainstorming were all considered secondary process goals. Students were introduced to process goals using an airplane activity consisting of four stations that had different models of airplanes. Following this activity, we introduced design constraints for the students to consider when constructing their models and split the students into teams of two to three. Throughout the investigation each team interacted with one or two of the PDP participants who acted as a facilitor to monitor and support each team's progress. We found that primary process and content goals were met when evaluated using both formative assessment and post-activity evaluation forms. However we encountered significant challenges during facilitation, which included problems with attendance, team dynamics, and disparities in communication and presentation styles and background knowledge. There were also challenges in the design and implementation of summative assessment tools for an engineering design activity. Overall, the CCD Modeling Activity was successful and well-received by the learners. Their experience with the engineering design process and knowledge about CCDs is highly relevant to potential future careers in the Hawai'i high-tech industry.

\section{Goals for Learners}
Since this activity fits into the ETRO 102 course, it is designed to help learners understand the fundamentals of how a Charge Coupled Device works. There are some fundamental concepts that must be taught to address what a CCD does, how a CCD functions, as well as some of the technical issues that arise in the use of CCDs.

At a basic level, students should understand that a CCD is made of material that converts a packet of light in to an electron. This material is structured in a regular array of pixels and that there is some process that moves electrons around within pixels. These electrons are trapped within the material of an individual pixel until an external process called ``readout'' is performed. Once this simplified concept of a CCD is understood second-tier content goals such as pixel well capacity, analog-to-digital conversion and bit depth, readout speed, readout method, charge-transfer-efficiency (CTE) and various noise sources can be addressed.

One major cognitive process we wished to introduce was the creation and use of a variety of models in the engineering process. Models can span a wide range of types from small and inexpensive illustrations to full-scale prototypes.  There are various trade-offs associated with creating different models and understanding trade-offs was part of this cognitive process goal. The primary process goal was to have learners understand modeling by being able to create and describe models. Students should also understand how modeling fits into the overall engineering design process. Secondary process goals included blueprinting, prototyping, validation, communication, and brainstorming. These content and process goals were chosen to fit in the context of the ETRO 102 course because many graduates of the program will pursue service and maintenance careers at the various remote-sensing facilities in Hawai'i.

\begin{table}[!ht]
\begin{center}
\caption{Activity Schedule}
\smallskip

{\small
\begin{tabular}{ccc}
\tableline
\noalign{\smallskip}
&Section & Time \\
\noalign{\smallskip}
\tableline
\noalign{\smallskip}
Day 1 & Intro to Activity \& Modeling & 10 min\\
&Airplanes Activity & 15 min\\
&Intro to CCDs \& Blueprints & 20 min\\
&Blueprinting & 60 min\\
&Presenting Blueprints & 30 min\\
\tableline
\noalign{\smallskip}
Day 2 &Model Building & 75 min\\
&Model Presentations & 30 min\\ 
&Synthesis \& Conclusion & 30 min\\
\noalign{\smallskip}
\tableline
\end{tabular}
}
\end{center}
\end{table}

\section{Activity Description}
There were a total of five hours in two lab sections allotted to this activity. Following an introduction to the activity we proceeded to a starter activity using airplane models that was designed to help the learners understand the trade-offs involved in engineering modeling. The starter was also designed to act as a scaffold to encourage the learners to consider the pros and cons of their design of a model demonstrating a CCD's readout mechanism.  After this more general introduction to modeling was complete, we presented the background motivation for CCDs.  The remainder of the first day was spent introducing the concept of blueprinting, and having the learners design and present their blueprints for their CCD model. The second day of class involved the learners building and presenting models of CCDs, as well as a final synthesis lecture. The activity timeline is presented in Table 1.

\subsection{Airplanes Activity}
\begin{wrapfigure}{r}{0.6\textwidth}
\begin{center}
\includegraphics[height=0.65\textwidth, angle=0]{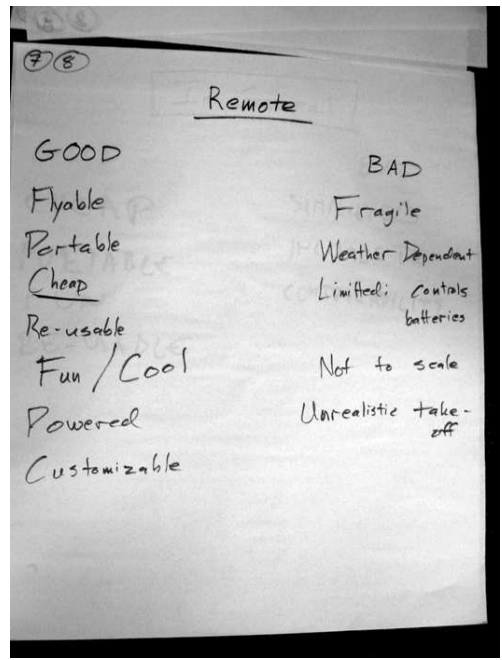}
\end{center}
\caption{List of pros and cons of the remote-controlled airplane model.  Most, but not all, of the listed characteristics were pointed out by the students. The underlined characteristic (``cheap'') was found to be common to several models. The common characteristics were planned for and summarized by the facilitating team.}
\end{wrapfigure}
This activity was intended to provide a hands-on approach to models and their importance in engineering design. Students inspected different airplane models (die cast, remote-controlled, computer simulator, balsa wood) and compiled a list of aspects of each model that represented, or failed to represent, a real airplane. Ultimately they shared their findings with the rest of the class.  Following the first day of the lab, the airplanes activity received positive feedback from the students in 2008 and 2009.  In addition to introducing the concept of modeling in engineering, this section served several secondary purposes. It allowed us to group the students into teams (of two or three), and gave them a chance to interact with each other, the facilitating team, and the workspace (laboratory).  The airplanes activity was meant to both relax the students and focus them on the lab, team work, and constructive and critical thinking common to engineers. It was also at this point that certain engineering concepts were first explored, and which would be revisited and practiced by the students later during the CCD modeling activity. These concepts are all part of the engineering design process and include the use of models, the necessity of trade-offs in models and engineering in general, and inter- and intra-team communication.

The college students were certainly familiar with airplanes to at least the level necessary to participate in the activity. The familiarity of the objects to be inspected was in turn intended to generate interest in the lab as a whole, as well as introduce some engineering concepts and practices in a setting that the students would find more comfortable than one in which we dove straight into more demanding CCD-related content. Finally, the idea of an airplane easily transcends all cultures and genders, so no objections stemming from the students' cultural, technical or other backgrounds were expected.

After a short introduction to the airplanes activity, each member of the facilitation team stood by one type of the airplane models.  Students were given a chance to examine the models closely, and the facilitators were present to ensure safety and to briefly demonstrate the operation of the more complex models, most notably the computer simulator. In addition, the facilitators unobtrusively observed and guided the discussion among the students. Each student team had a chance to look at each of the models, while being encouraged to note the trade-offs each model represented.

Next, the facilitators asked the students to name the pros and cons of each model. One of the facilitators moderated the discussion while another wrote the relevant properties of the models on a poster sheet, an example is shown in Figure 1. Finally, the facilitators reemphasized the main points with a short synthesis presentation, which in addition to modeling in engineering included the importance of critical thinking and effective communication of ideas in engineering. The students were thus better prepared to practice the same skills in earnest during the remainder of the lab. However, since airplanes are not related to CCDs in terms of content, the facilitators kept a close watch on the time spent on this lab component, even when it proved to be enjoyable to the students.

\subsection{Introduction to CCDs and Blueprints}

In a brief presentation CCDs were compared to photographic film and examples of their applications were reviewed. Then, we introduced blueprints as a crucial preliminary step in planning a model. At this point we revealed the task of constructing a CCD model, and explained that it should simulate as many previously presented functions and features of a CCD as realistically as possible.  In order to stimulate the students' ideas in constructing their models we showed them materials we had collected for their use. These materials were inexpensive items that the facilitators had thought of while brainstorming how they might build a CCD model.  For example, we had small items such as beads and push-pins that could represent photo-electrons in the CCD, and ice trays and plastic screens that could be used as the pixel array. We also included various household and craft items, such as rubber bands, dowels, and washers that might not have a use that was obvious to us, but could conceivably be important in a design.  In addition, we offered to bring to the next class any materials that the students thought they would like to use, provided that they were not too expensive.

\subsection{Blueprinting}

In practice, it is impossible to model all aspects of a CCD using only household materials in a few hours time.  The key issue that the students were able to address within these constraints is the shift-register mechanism of CCD readout, in which varying voltage on the pixel array shifts electrons along rows and columns of the array.   We did not present the shift-register mechanism to the students at this point, but noted that a CCD is able to readout the amount of charge in a particular pixel without having physical access to it.  In addition to demonstrating the readout concept, the student teams were expected to note the unavoidable trade-offs their model would realize. Finally, the blueprint served as the main tool for communicating ideas within the team during the planning, as well as during a blueprint presentation to the class at the end of the first day.

\begin{wrapfigure}{r}{0.6\textwidth}\begin{center}
\includegraphics[height=0.65\textwidth, angle=0]{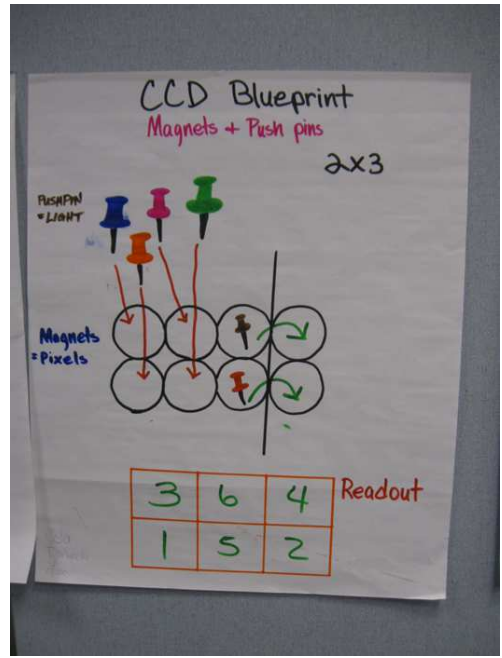}\end{center}
\caption{A student team's CCD model blueprint, demonstrating the simulation of CCD pixels and the readout mechanism. First-day blueprints were allowed and encouraged to be modified during the Model Building phase.}
\end{wrapfigure}

This activity was the most demanding portion of the first day and was consequently allotted the largest amount of time. The students concentrated on planning their CCD model, working in groups of two to three. The blueprinting materials provided in the lab were large white sheets and writing and drawing tools. A finished CCD blueprint is shown on Figure 2. Unlike in the airplanes activity, the facilitators were not constantly present at any student team's station, but instead shadowed the teams, by periodically checking the progress and providing guidance as needed. Each facilitator was assigned to one or more teams. During the presentations of the blueprints, the students were encouraged, by example, to ask questions and point out any interesting details about the planned models. Following the presentations, the facilitating team evaluated the first day's success and subsequently planned for the second day.

\subsection{Model Building}

The second day of the activity started with a brief review of the previous session and then learners started building their models. They could use simple office materials for this purpose. They were told to follow their blueprints from the previous day, keeping in mind the constraints about the readout process. In cases where the blueprints were not feasible, the facilitators provided help to guide students through alternate routes as delicately as possible in order for the the learners to maintain ownership of the design. One facilitator was assigned to two or three groups of students. During the 75 minutes dedicated to this section, facilitators kept track of the students' progress by checking in on them every 10 to 15 minutes. We decided not to have facilitators rotate between groups because of the added time that would be needed for the learners to repeatedly explain their model design. We did however, have a checkpoint in the middle of the activity where all facilitators got together to exchange their experience and advice for certain situations in their groups. Several groups were very quick in modeling the readout as desired, so their facilitator asked about extra features or issues, such as dead pixels, that the team could implement in their models.  A completed model build by one of the teams is shown in Figure 3.

\begin{wrapfigure}{r}{0.6\textwidth}
\begin{center}
\includegraphics[height=0.4\textwidth, angle=0]{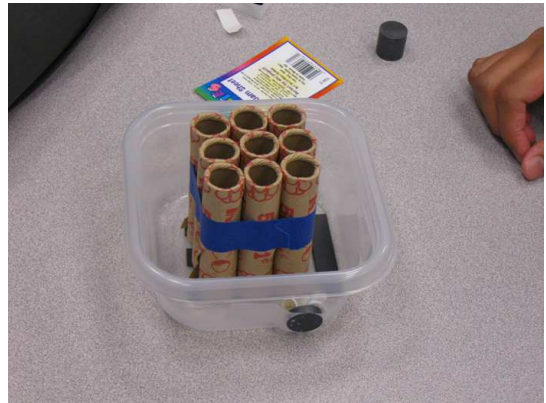}
\end{center}
\caption{A student team's early CCD model. The paper tubes representing catch pins or beads, which stick to the magnetic strips at the bottom of the cup. The pins and beads can be transferred to the magnetic strip on the right, thus (partially) simulating the readout.}
\end{wrapfigure}

\subsection{Student Presentations}
Following model building the students had time to prepare a short presentation demonstrating their models and explaining the models' strengths and weaknesses.  For example, one group came up with a 2D array of paper cups (as pixels) and beads (as photons). They then simulated the readout by pouring the contents of a cup into the cup next to it in the same row. Each group discussed the limitations of their models and as well as which CCD artifacts it could simulate. Most groups where able to converge to the concept of readout based on the introduction that they were given. The groups were able to practice presentation process skills, some of which had been briefly introduced during the activity and included instructions that everyone in the group should talk, they should keep their explanations concise, and that drawings and figures are encouraged.

\subsection{Synthesis}
The classroom experience concluded with a synthesis talk that lasted about 30 minutes.  The talk covered all content goals of the activity and some of the engineering process goals. We started with the history of the CCD and mentioned that October 2009 (when the activity was taught) was the 40th anniversary of its invention and that this innovation was awarded with the 2009 Nobel Prize in Physics. Copies of the original hand writings of the CCD concept were shown to emphasize how important lab notebooks are in research. We went through how CCDs work, the building blocks, the readout mechanism and simplified models to exemplify the readout. Since each of the facilitators uses CCDs in their daily research work, we decided that it would be best to have each facilitator talk about applications of CCDs in their field. The synthesis talk ended with a reminder of the engineering design cycle, which is described by \citet{masseng}.

\section{Facilitation}
A number of facilitation techniques were implemented during the course of the activity.  For example, during the blueprinting and building phases, the facilitator's role was to make sure the students followed the guidelines of a good blueprint and worked within the specified constraints of a CCD. At the same time, they monitored students' progress towards grasping the major content goal of the activity, which was the CCD readout process. Facilitators would intervene periodically by asking questions about the group's ideas of the readout process every 10 to 15 minutes.  If necessary the facilitator could challenge the students' ideas as described by their blueprints or implemented in their models and ask for clarification about how the design fit within constraints and trade-offs of CCD readout.

\subsection{Facilitation Challenges}
Some of the challenges we faced during facilitation were: a dominating team member, students who were reluctant to contribute, and students who arrived late for class.  For example, a few teams had a dominating team member who thought he was on the right path to demonstrating the readout process (but was not necessarily), and was thus preventing the other team members from experimenting and contributing. In several groups, participants had problems being assertive, this was particularly common among female students who were paired with a male.  In both cases the main facilitation strategy was to play down the role of the dominant member by interacting directly with the less vocal member of the team, and if necessary pulling the dominant member aside and asking that he make more of an effort to let the other members contribute. Another issue with that came up in this activity was the problem of what to do with students who arrived late, or missed the first day. We had a teaching consultant who was observing our activity, and we asked that he give the tardy students a brief introduction to the activity and then help them integrate into one of the design teams.
 
Another challenge we encountered was dealing with a large range of students' background knowledge. Some students got bored since they had been exposed to the material before. Having tiered content goals helped the facilitators be prepared to bring up more challenging cases for students with more advanced backgrounds. For example, the main content goal was to explain the readout process and the second tier goal was to model bad pixels, color and read noise in a CCD. Any team that could demonstrate mastery of a first-tier goal was guided towards a second-tier goal for further investigations. The idea was to have various entry points so that all students could participate.

\section{Assessment Plan}
During the course of the activity the facilitators interactions with the students helped to assess the prior knowledge and progress toward grasping the content and process goals of the activity.  This formative assessment helped to guide future interactions with the students.   At the end of the activity we expected the students to be able to describe the readout process through their models. In order to assist our evaluation of the students' explanations of the readout process, we wrote a rubric based on the Claim-Evidence-Reasoning framework described by \citet{mcn}.  We adapted this rubric template to fit the engineering context of our activity.  One issue we encountered was that facilitators who  assessed the presentations of a group of students whom they had not worked with usually gave a lower score than the facilitator who had interacted with the group throughout the activity. We decided that averaging the grades would solve this problem. It was also difficult to assess a students' understanding of a readout process if they had problems presenting their work.

\section{Future Considerations}
The formative and summative assessment process could be further refined, particularly with regard to evaluations of the student presentations.  This would involve generating a more developed rubric that clearly specifies evidence that the students have met both content and process goals.  Another option would be to try to apply the assessment criteria more broadly, such as within the activity itself, and possibly with a written assignment.  Additionally, issues related to group dynamics could be considered in more depth in the future.

\section{Summary}
We designed an activity to introduce CCDs as part of the ETRO 102 course at MCC. This activity was conceived and initially implemented in 2008. The main goal was to help students understand how CCDs work; most importantly, to understand the readout process in a CCD. The process goals included having the students understand modeling in engineering, particularly the importance of blueprints.  With these goals in mind we presented some material in a traditional lecture format, but also had the students design, build and present a model of a CCD that focused on the CCDs readout mechanism.  Though we did encounter some challenges in the implementation of this activity, we found that overall it went very smoothly and we had very positive feedback from the students.

\acknowledgments
Bradley Jacobs, Nicholas Moskovitz, Isar Mostafanezhad, Elisabeth Reader, Ciril Rozic, Ryan Swindle, and Johnny Tam participated on the PDP design teams.  They were assisted by the ISEE staff, with particular support from David Harrington, Lisa Hunter, Lani Lebron, and Barry Kluger-Bell.  ETRO 102 was taught by Mark Hoffman and Elisabeth Reader. We used facilities at Maui Community College, Kahului, and course materials were provided through the Akami Maui Workforce Initiative. The authors acknowledge the National Science Foundation Science and Technology Center funding of the Center for Adaptive Optics, managed by the University of California, Santa Cruz, under cooperative agreement No.~AST-9876783.

\bibliography{mostafanezhad}

\begin{thebibliography}{}
\expandafter\ifx\csname natexlab\endcsname\relax\def\natexlab#1{#1}\fi
\expandafter\ifx\csname url\endcsname\relax
  \def\url#1{\texttt{#1}}\fi
\expandafter\ifx\csname urlprefix\endcsname\relax\def\urlprefix{URL }\fi
\providecommand{\eprint}[2][]{\url{#2}}

\bibitem[{Driscoll et~al.(2006)Driscoll, Peyser, Reale, Anderson, Chernow,
  Plummer, Schaefer, Thernstrom, Thomas, \& Fredrick}]{masseng}
Driscoll, D.~P., Peyser, J.~A., Reale, A., Anderson, C., Chernow, H., Plummer,
  P., Schaefer, R.~R., Thernstrom, A.~M., Thomas, H.~M., \& Fredrick, T. 2006,
  Massachusetts Science and Technology/Engineering Curriculum Framework,
  Massachusetts Department of Education.
  \urlprefix\url{http://www.doe.mass.edu/frameworks/scitech/1006.pdf}

\bibitem[{Hunter et~al.(2008)Hunter, Metevier, Seagroves, Porter, Raschke,
  Kluger-Bell, Brown, Jonsson, \& Ash}]{PDPdesc}
Hunter, L., Metevier, A., Seagroves, S., Porter, J., Raschke, L., Kluger-Bell,
  B., Brown, C., Jonsson, P., \& Ash, D. 2008, Cultivating Scientist- and
  Engineer-Educators: The CfAO Professional Development Program.
  \urlprefix\url{http://isee.ucsc.edu/participants/programs/CfAO_Prof_Dev_Prog%
ram.pdf}

\bibitem[{McNeill \& Krajcik(2009)}]{mcn}
McNeill, K.~L., \& Krajcik, J. 2009, Journal of the Learning Sciences, 18, 416

\end{thebibliography}

%\begin{thebibliography} 
%\bibitem[Driscoll et al.(2006)]{masseng} Driscoll, D. P., et al. 2006, http://www.doe.mass.edu/frameworks/scitech/1006.pdf.
%\bibitem[Hunter et al.(2008)]{PDPdesc}	Hunter, L., Metevier, A., Seagroves, S., Porter, J., Raschke, L., Kluger-Bell, B., Brown, C., Jonsson, P., Ash, D.  2008, http://isee.ucsc.edu/participants/programs/CfAO\_Prof\_Dev\_Program.pdf.
%\bibitem[McNeill \& Krajcik(2009)]{mcn} McNeill, K. L., Krajcik, J. 2009, Journal of the Learning Sciences, 18:3, 416.
%\bibitem[Morzinski et al.(2010a)]{morza} Morzinski, K., Crockett, C., Crossfield, I. 2010, This Proceedings.
%\bibitem[Morzinski et al.(2010b)]{morzb} Morzinski, K., Azucena, O., Downs, C., Favoloro, T., Park, J., U, V. 2010, This Proceedings. 
%\end{thebibliography}

\end{document}